\documentclass[aps,pra,reprint,amsmath,amssymb,superscriptaddress,longbibliography]{revtex4-2}

\usepackage{bm}
\usepackage{mathtools}
\usepackage{booktabs}
\usepackage[colorlinks=true,allcolors=blue]{hyperref}
\usepackage{amsthm}
\usepackage{braket}

\newcommand{\cE}{\mathcal{E}}

\newcommand{\id}{\operatorname{id}}
\newcommand{\conv}{\operatorname{conv}}

\newcommand{\Iacc}{I_{\mathrm{acc}}}

\newtheorem{theorem}{Theorem}
\newtheorem{proposition}[theorem]{Proposition}
\newtheorem{lemma}[theorem]{Lemma}
\newtheorem{definition}[theorem]{Definition}

\begin{document}

\title{Holevo-Bounded Total Information from Pair Distributions of
Alternative Quantum Measurements}

\author{Baichu Yu}
\email{yubaichu@mail.ustc.edu.cn}
\affiliation{Quantum Science Center of Guangdong-Hong Kong-Macao Greater Bay Area, Shenzhen 518045, China}

\author{Zhang Zhou}
\affiliation{Institute of Automation, Chinese Academy of Sciences, Beijing 100190, China}

\author{Zhiyuan Li}
\affiliation{CAS Zidong Taichu Co., Ltd., Beijing 100190, China}

\author{Jiajun Zhang}
\affiliation{Institute of Automation, Chinese Academy of Sciences, Beijing 100190, China}
\affiliation{CAS Zidong Taichu Co., Ltd., Beijing 100190, China}

\date{November 9, 2026}

\begin{abstract}
The accessible information of a quantum ensemble can be strictly smaller than
its Holevo information because a single-copy measurement may discard part of the
label--system correlation. We ask whether information represented across two
alternative measurement contexts can bridge this gap without assigning a
joint distribution to their outcomes. Building on the relative-entropy-based projection method introduced by Harder et al., we define the directed posterior novelty of each record relative to the other, and an associated total-information quantity. For posterior-orthogonal records,
this quantity equals the sum of the two record mutual informations. 
For
finite-dimensional classical--quantum ensembles, we prove that optimizing this quantity over pairs of POVMs yields
a value between the accessible and Holevo information. The upper bound follows by showing that the directed posterior novelty provides a lower bound on the infimum of the measured-relative-entropy recovery error.
The uniform BB84 ensemble provides an example in which the quantity saturates
the Holevo information while strictly exceeding the accessible information.
Finally, we analytically establish the quantum validity of a
redundancy-based information causality condition, resolving a question
that has remained open since the condition was introduced.
\end{abstract}

\maketitle

\section{Introduction}
\label{sec:introduction}
A quantum ensemble may encode more classical information than can be
accessed by any single measurement.  For an ensemble
$\mathcal E=\{p_a,\rho_a\}_a$, Holevo's theorem establishes
\begin{equation}
 I_{\mathrm{acc}}(\mathcal E)\leq \chi(\mathcal E),
 \label{eq:intro-holevo}
\end{equation}
where $I_{\mathrm{acc}}(\mathcal E)$ is the largest mutual information
obtainable from a single measurement, while $\chi(\mathcal E)$ is the Holevo
quantity and equals the mutual information of the classical-quantum (cq) state associated with $\mathcal E$ before measurement.  The
inequality in Eq.~\eqref{eq:intro-holevo} can be strict: a single-copy
measurement may irreversibly lose part of the label--system correlation
\cite{holevo1973bounds}.

Several related approaches clarify how the Holevo information may be attained or
how measurement-induced information loss may be characterized.  Quantum
channel-coding theorems show that $\chi$ is attainable asymptotically as a
communication rate by encoding into long blocks and using collective or
sequential decoding measurements
\cite{hausladen1996classical,schumacher1997sending,holevo1998capacity}.
Other work strengthens measurement-specific information bounds by subtracting
terms associated with information retained in the conditional postmeasurement
ensembles \cite{schumacher1996limitation,jacobs2006bound}; 
later formulations distinguish information extracted by the
measurement, left accessible in the postmeasurement system, or destroyed
\cite{han2018residual}.

In this paper, we ask whether statistics from alternative single-copy
measurement contexts can represent more of the ensemble information than any
one context while remaining bounded by $\chi(\mathcal E)$.  Specifically, we
seek an information quantity determined solely by the observed label--record
pair distributions obtained by measuring individual systems prepared
according to the ensemble.

We consider the simplest setting of two alternative
measurements.  Even in this case, the question is subtle because their
outcomes need not admit a physically meaningful joint distribution,
particularly when the measurements are incompatible.  The information
represented across the two measurements therefore cannot in general be quantified
as the mutual information of a
joint outcome record.  The
available data are instead the pair distributions $p(a,b_1)$ and
$p(a,b_2)$ between the label $A$ and the outcome of each
measurement.  
A suitable total-information quantity should therefore
be determined by these two pair distributions alone.

Partial information decomposition (PID) \cite{williams2010nonnegative,harder2013bivariate,griffith2014intersection,griffith2015quantifying,ince2017measuring,finn2018pointwise,kolchinsky2022novel} provides a language for separating
the redundant, unique, and synergistic information that several records carry
about a common target.  Its conventional bivariate formulation, however,
begins with a joint distribution $p(a,b_1,b_2)$.  When only
$p(a,b_1)$ and $p(a,b_2)$ are specified, neither the joint-record mutual
information nor a synergistic contribution is fixed.  The
Harder--Salge--Polani (HSP) projected-redundancy construction is particularly
suited to this setting because its bivariate redundancy depends only on the
two label--record pair distributions \cite{harder2013bivariate}.

Motivated by the HSP projected redundancy, we introduce the directed posterior novelty $\delta_{j\to i}$, which
quantifies posterior structure generated by $B_j$ that cannot be represented
by probabilistic mixtures of posteriors generated by $B_i$.  Using the two
directed novelties, we define the fixed-pair pair-distribution (PD) total information as
\begin{multline}
 R_{\mathrm{PD}}(A;B_1,B_2):=\max\!\bigl\{
 I(A;B_1)+\delta_{2\to1},\\
 I(A;B_2)+\delta_{1\to2}\bigr\}.
 \label{eq:intro-fixed-hsp}
\end{multline}
Each branch combines the information represented in one context with the
posterior structure generated in the other that is not representable in the first. 
The term “total” refers here to the information structure represented collectively by the two specified pair distributions. It does not describe information jointly obtainable by applying both measurements to a single copy.
In fact, this novelty-based definition is equivalently
expressed in terms of the HSP projected redundancy \cite{harder2013bivariate} as
\begin{equation}
 R_{\mathrm{PD}}(A;B_1,B_2)
 =
 I(A;B_1)+I(A;B_2)
 -I_{\cap}^{\mathrm{HSP}}(A;B_1,B_2).
 \label{eq:intro-hsp-redundancy-form}
\end{equation}
We establish three basic properties of \(R_{\mathrm{PD}}\): it lower-bounds
the joint mutual information of every compatible coupling, lies between the
larger individual mutual information and the sum of the two mutual
informations, and reduces to this sum when the two records are posterior-orthogonal.

We then apply this construction to two alternative POVMs on a
finite-dimensional cq ensemble $\mathcal{E}$.  
Optimizing the fixed-pair quantity over
POVM pairs defines $R_{\mathrm{PD}}(\mathcal E)$, which is the PD total information for the ensemble $\mathcal E$.  Our main result is the
hierarchy
\begin{equation}
 I_{\mathrm{acc}}(\mathcal E)
 \leq R_{\mathrm{PD}}(\mathcal E)
 \leq \chi(\mathcal E).
 \label{MainHierarchy0}
\end{equation}
The lower bound follows because a single measurement is included by adjoining
a trivial second measurement.  For the upper bound, we prove the directed
estimate
$\delta_{j\to i}\leq I(A:H)-I(A;B_i)$, where $H$ is the premeasurement
quantum system.  
The proof shows that the directed posterior novelty lower-bounds the infimum of the measured-relative-entropy recovery error over all recovery maps, which is in turn upper-bounded by the conditional mutual information associated with the measurement dilation.
As an example, we show that, for the uniform BB84 ensemble
\cite{bennett2014quantum}, the PD total information of the
posterior-orthogonal Pauli-\(Z\) and Pauli-\(X\) records strictly exceeds the
accessible information and saturates the Holevo bound.

Finally, we apply the hierarchy in Eq.~\eqref{MainHierarchy0} to the
HSP redundancy-based information causality condition proposed in
Ref.~\cite{yu2026information}. Although numerical evidence
supported its validity for quantum correlations, whether the condition
is satisfied by every quantum realization has remained open. We resolve this question analytically under the
corresponding protocol assumptions, thereby establishing the condition
as a quantum-valid formulation of information causality. Consequently,
its violation certifies beyond-quantum correlations within this setting.

A related direction is the extension of PID to quantum systems, where
information in quantum states or ensembles, including the Holevo information,
is decomposed into distinct informational components
\cite{van2023quantum,ericson2026defining,nico2026quantum}.  
The present
construction instead concerns alternative measurement records and is defined
directly from their two pair distributions with the label,
without assuming a joint outcome distribution.  It therefore provides a
measurement-context notion of total information rather than a full quantum
PID.

\section{Total information from two pair distributions}
\label{sec:hsp}

Let $A$ be a finite random variable with prior distribution $p(a)$. Suppose two observations provide pair distributions
\begin{equation}
 p_i(a,b_i),\qquad i\in\{1,2\},
\label{eq:pair-marginals}
\end{equation}
with the same $A$ marginal. No joint distribution of $(A,B_1,B_2)$ is assumed. 
Let all logarithms take base two, and let $D(p\|q)$ denote classical relative entropy.

For each $B_i$, define the posterior polytope
\begin{equation}
 \mathcal C_i=\conv\{p_i(A|b_i):p_i(b_i)>0\}.
\label{eq:posterior-polytope}
\end{equation}

For $i\neq j$, introduce the directed posterior novelty of $B_j$ relative to $B_i$ as
\begin{equation}
 \delta_{j\to i}
 =\sum_{b_j}p_j(b_j)\inf_{r\in\mathcal C_i}
 D\!\left(p_j(A|b_j)\middle\|r\right).
\label{eq:delta-def}
\end{equation}
The novelty is the average minimum divergence between the posterior distributions generated by $B_j$ and the posterior polytope generated by $B_i$. 
It measures the posterior structure $B_j$ contains that cannot be reproduced by probabilistic mixtures of the posteriors generated by $B_i$.
Combining the information represented by one record with the directed novelty
contributed by the other, we define the fixed-pair total information
\(R_{\mathrm{PD}}(A;B_1,B_2)\) as in
Eq.~\eqref{eq:intro-fixed-hsp}. When the pair of records is clear or not
relevant to the discussion, we abbreviate this quantity as
\(R_{\mathrm{PD}}\).

The quantity $R_{\mathrm{PD}}$ can be defined alternatively through the HSP redundant information defined in Ref.~\cite{harder2013bivariate}. The HSP projected information from $B_{j}$ to $B_{i}$ is
\begin{align}
 I_A^{\pi}(B_j\searrow B_i)
 &=\sum_{b_j}p_j(b_j)
 \left[D\!\left(p_j(A|b_j)\middle\|p(A)\right)
 \right.\nonumber\\[-1mm]
 &\hspace{29mm}\left.
 -\inf_{r\in\mathcal C_i}D\!\left(p_j(A|b_j)\middle\|r\right)\right]
 \nonumber\\
 &=I(A;B_j)-\delta_{j\to i}.
\label{eq:projected-information}
\end{align}
The HSP redundant information is defined as the smaller of the two directed projected informations
\begin{equation}
 I_{\cap}^{\mathrm{HSP}}(A;B_1,B_2)
 =\min\{I_A^{\pi}(B_1\searrow B_2),
          I_A^{\pi}(B_2\searrow B_1)\}.
\label{eq:hsp-redundancy}
\end{equation}
Subtracting this redundancy from the sum of the two individual mutual informations gives
\begin{align}
&I(A;B_1)+I(A;B_2)
   -I_{\cap}^{\mathrm{HSP}}(A;B_1,B_2)
\nonumber\\
 =&\max\{I(A;B_2)+\delta_{1\to2},
          I(A;B_1)+\delta_{2\to1}\}\nonumber\\
 =&R_{\mathrm{PD}}.
\label{eq:total-max-form}
\end{align}
Equation~\eqref{eq:total-max-form} expresses $R_{\mathrm{PD}}$ through the picture of PID, showing that the quantity includes the redundant contribution and the contributions represented uniquely across the two posterior families, without assigning a synergistic component. 

We present three properties of $R_{\mathrm{PD}}$. First, although \(R_{\mathrm{PD}}\) is defined without a
joint distribution, it lower-bounds the joint mutual information of every
coupling consistent with the specified pair distributions. Let
\begin{equation}
 \Gamma(p_1,p_2)
 :=
 \left\{
 q_{AB_1B_2}:
 q_{AB_1}=p_1,\;
 q_{AB_2}=p_2
 \right\}.
 \label{eq:coupling-set}
\end{equation}
Lemma~7 of Ref.~
\cite{harder2013bivariate} implies
\begin{equation}
 R_{\mathrm{PD}}(A;B_1,B_2)
 \leq
 \min_{q\in\Gamma(p_1,p_2)}
 I_q(A;B_1,B_2).
 \label{eq:range-joint}
\end{equation}
The couplings in \(\Gamma(p_1,p_2)\) are mathematical extensions of the two
pair distributions and need not represent physically simultaneous records.

Second, because \(p(A)\in\mathcal C_i\), the directed novelty satisfies
\begin{equation}
 0\leq\delta_{j\to i}\leq I(A;B_j).
\end{equation}
Consequently,
\begin{equation}
 \max\{I(A;B_1),I(A;B_2)\}
 \leq R_{\mathrm{PD}}
 \leq I(A;B_1)+I(A;B_2).
 \label{rangeHSP}
\end{equation}

The third property concerns the following class of records. 
\begin{definition}[posterior-orthogonal records]
Assuming \(p(a)>0\) on its
support, call \(B_1\) and \(B_2\) posterior-orthogonal about \(A\) if
\begin{equation}
 \sum_a
 \frac{
 [p_1(a|b_1)-p(a)]
 [p_2(a|b_2)-p(a)]
 }{p(a)}
 =0
 \label{posterior-orthogonality}
\end{equation}
for every \(b_1,b_2\) of positive probability.  
\end{definition}
\begin{proposition}\label{propOrth}
For posterior-orthogonal records, the directed novelty of either record
relative to the other equals its full mutual information with \(A\):
\begin{equation}
 \delta_{j\to i}=I(A;B_j),
 \qquad i\neq j.
 \label{eq:orthogonal-full-novelty}
\end{equation}
Consequently
\begin{equation}
 R_{\mathrm{PD}}(A;B_1,B_2)
 =
 I(A;B_1)+I(A;B_2).
 \label{eq:orthogonal-PD}
\end{equation}
\end{proposition}
Thus, the upper bound in Eq.~\eqref{rangeHSP} is attained. The proof is
given in Appendix~\ref{app:posterior-orthogonality}.

\section{Total information of a quantum ensemble}
\label{sec:quantum-ensemble}

\subsection{Classical--quantum ensembles and measurements}

Let
\begin{equation}
 \mathcal E=\{p_a,\rho_a^H\}_{a\in A}
 \label{eq:ensemble-sec3}
\end{equation}
be a finite ensemble of quantum states on a finite-dimensional system $H$. Its average state is
\begin{equation}
 \bar\rho^H=\sum_a p_a\rho_a^H.
 \label{eq:average-state-sec3}
\end{equation}
The Holevo information of the ensemble is
\begin{equation}
 \chi(\mathcal E)
 =S(\bar\rho)-\sum_a p_a S(\rho_a),
 \label{eq:holevo-sec3}
\end{equation}
where $S(\cdot)$ denotes the von Neumann entropy.
$\chi(\mathcal E)$ quantifies the total correlation between the
classical label and the unmeasured quantum system.

The ensemble $\mathcal E$ can be equivalently represented
by a classical--quantum (cq) state
\begin{equation}
 \rho^{AH}=\sum_{a\in\mathcal A}p_a
 |a\rangle\!\langle a|^A\otimes\rho_a^H,
 \label{eq:cq-state-proof}
\end{equation}
for which
\begin{equation}
I(A:H)_\rho
 =S(\rho^A)+S(\rho^H)-S(\rho^{AH})= \chi(\mathcal E).
 \label{eq:chi-qmi-proof}
\end{equation}

A finite-outcome measurement $\mathsf M$ on $H$ is described by a
positive-operator-valued measure (POVM)
\begin{equation}
 \mathsf M=\{E_{b}\}_{b\in B},
 \qquad
 E_{b}\geq0,
 \qquad
 \sum_{b}E_{b}=\mathbb I_H.
 \label{eq:povm-sec3}
\end{equation}
Measuring system $H$ by $\mathsf M$ produces the
classical pair distribution
\begin{equation}
 p(a,b)
 =p_a\operatorname{Tr}\!\left(\rho_a E_{b}\right).
 \label{eq:measurement-pair-sec3}
\end{equation}
% The outcome probability and posterior distribution of the label are
% therefore
% \begin{align}
%  p_i(b_i)
%  &=\sum_a p_a\operatorname{Tr}\!\left(\rho_a E_{b_i}^{(i)}\right),
%  \label{eq:outcome-prob-sec3}\\
%  p_i(a|b_i)
%  &=\frac{p_a\operatorname{Tr}\!\left(\rho_a E_{b_i}^{(i)}\right)}
%  {p_i(b_i)},
%  \qquad p_i(b_i)>0.
%  \label{eq:posterior-sec3}
% \end{align}
The accessible information
\begin{equation}
 \Iacc(\cE)=\sup_{\mathsf M} I(A;B)
\label{eq:iacc-intro}
\end{equation}
is the largest mutual information obtainable with one measurement.
By data processing,
\begin{equation}
 \Iacc(\cE)\leq I(A:H)_\rho=\chi(\mathcal E).
 \label{eq:data-processing-sec3}
\end{equation}

\subsection{Two alternative measurement contexts}

We now consider two POVMs $\mathsf M_1$ and $\mathsf M_2$ on the same
quantum system.  They are understood as \emph{alternative} measurement
settings: on each copy of the ensemble, one chooses either $\mathsf M_1$ or
$\mathsf M_2$.  The two experiments determine the pair distributions
$p_1(a,b_1)$ and $p_2(a,b_2)$ in Eq.~\eqref{eq:measurement-pair-sec3}, and
both have the same marginal $p(a)=p_a$. No assumption of joint measurability is made.

Following Sec.~\ref{sec:hsp}, a total information $R_{\mathrm{PD}}(\mathcal E;\mathsf M_1,\mathsf M_2)$ can be determined with the pair distributions generated by each fixed pair of POVMs $\mathsf M_1$ and $\mathsf M_2$.

The PD total information of the ensemble is obtained by optimizing over
pairs of finite-outcome POVMs:
\begin{equation}
 R_{\mathrm{PD}}(\mathcal E)
 =\sup_{\mathsf M_1,\mathsf M_2}
 R_{\mathrm{PD}}(\mathcal E;\mathsf M_1,\mathsf M_2).
 \label{DefBySup}
\end{equation}
Equation~\eqref{DefBySup} should not be interpreted as the
information extracted by jointly applying both measurements to one copy.
Rather, it quantifies the information represented across the statistics of
two alternative measurement contexts.  The underlying resource is still a
single copy of the quantum system in each experimental run, and
incompatible outcomes are never assigned a joint value.

\subsection{Main result}

The main result of this paper is that the total information $R_{\mathrm{PD}}$ is bounded by the two standard
information measures of the ensemble.

\begin{theorem}[HSP--Holevo hierarchy]
\label{thm:main-hierarchy-sec3}
For every finite-dimensional cq ensemble $\mathcal E$,
\begin{equation}
 I_{\mathrm{acc}}(\mathcal E)
 \leq R_{\mathrm{PD}}(\mathcal E)
 \leq \chi(\mathcal E).
 \label{MainHierarchy}
\end{equation}
\end{theorem}

The two inequalities in Theorem~\ref{thm:main-hierarchy-sec3} establish two properties of the PD total information.  The lower bound
$I_{\mathrm{acc}}(\mathcal E)\leq R_{\mathrm{PD}}(\mathcal E)$ shows that
allowing two alternative measurement contexts can reveal additional posterior structure
not captured by the best single measurement.  
The upper bound
$R_{\mathrm{PD}}(\mathcal E)\leq\chi(\mathcal E)$ shows that combining the
information of one measurement with the directed novelty of another
cannot yield more total information about the label $A$ than the
quantum system $H$ contains before measurement.

The proof of the lower bound is  immediate using the first inequality of Eq.~\eqref{rangeHSP} and the definition Eq.~\eqref{DefBySup}.  
The proof of the upper bound is the nontrivial part.  
Its key ingredient is the following
directed inequality:
\begin{equation}
 \delta_{j\to i}
 \leq I(A:H)_\rho-I(A;B_i),
 \qquad i\neq j.
 \label{novelty2CE}
\end{equation}
The right-hand side is the quantum mutual information lost when system $H$ is
measured by $\mathsf M_i$. 
Through a Stinespring isometry,  
this loss becomes a conditional mutual information involving
the measurement environment.  
A measured-relative-entropy recovery
theorem lower bounds this quantity by a recovery error \cite{brandao2015quantum}.  
Finally,
data processing and the classical
relative-entropy chain rule, in turn, bound this recovery error from below
by the projection cost $\delta_{j\to i}$.  
Thus the posterior novelty supplied by an
alternative measurement cannot exceed the information lost in the 
reference measurement. 
Adding $I(A;B_i)$ to both sides of \eqref{novelty2CE} gives
\begin{equation}
 I(A;B_i)+\delta_{j\to i}
 \leq I(A:H)_\rho
 =\chi(\mathcal E).
 \label{eq:branch-holevo-sec3}
\end{equation}
Taking the
supremum over all POVM pairs then gives
$R_{\mathrm{PD}}(\mathcal E)\leq\chi(\mathcal E)$.
The full proof for theorem \ref{thm:main-hierarchy-sec3} is given in
Sec.~\ref{sec:proof-main}.

\subsection{Strict separation and saturation of the Holevo bound}

The hierarchy in Eq.~\eqref{MainHierarchy} permits
\(R_{\mathrm{PD}}\) to exceed the accessible information while remaining
bounded by the Holevo information. The uniform BB84 ensemble provides an
analytic example in which the lower inequality is strict and the upper
inequality is saturated. Consider
\begin{equation}
 \mathcal E_{\mathrm{BB84}}
 =
 \left\{
 \frac14,|0\rangle;
 \frac14,|1\rangle;
 \frac14,|+\rangle;
 \frac14,|-\rangle
 \right\},
 \label{eq:BB84-ensemble}
\end{equation}
where
\begin{equation}
 |\pm\rangle=\frac{|0\rangle\pm|1\rangle}{\sqrt2}.
\end{equation}
Since the signal states are pure and their average state is maximally mixed,
\begin{equation}
 \chi(\mathcal E_{\mathrm{BB84}})=1.
 \label{eq:BB84-Holevo}
\end{equation}

Let \(B_Z\) and \(B_X\) be the records generated by measurements in the
Pauli-\(Z\) and Pauli-\(X\) bases, respectively. Each measurement perfectly
distinguishes the two states in its own basis, while the states in the other
basis produce uniformly random outcomes. Hence
\begin{equation}
 I(A;B_Z)=I(A;B_X)=\frac12.
 \label{eq:BB84-mutual-information}
\end{equation}
A direct calculation from the four posterior distributions shows that the
\(Z\)- and \(X\)-measurement records satisfy
Eq.~\eqref{posterior-orthogonality}. They are therefore
posterior-orthogonal about \(A\), and
Proposition~\ref{propOrth} gives
\begin{equation}
 \delta_{X\to Z}=I(A;B_X)=\frac12,
 \qquad
 \delta_{Z\to X}=I(A;B_Z)=\frac12.
\end{equation} 
and consequently
\begin{equation}
 R_{\mathrm{PD}}(\mathcal E_{\mathrm{BB84}};Z,X)
 =
 I(A;B_Z)+I(A;B_X)
 =1.
 \label{eq:BB84-fixed-PD}
\end{equation}
Together with Theorem~\ref{thm:main-hierarchy-sec3}, this yields
\begin{equation}
 1
 =
 R_{\mathrm{PD}}(\mathcal E_{\mathrm{BB84}};Z,X)
 \leq
 R_{\mathrm{PD}}(\mathcal E_{\mathrm{BB84}})
 \leq
 \chi(\mathcal E_{\mathrm{BB84}})
 =1,
\end{equation}
and hence
\begin{equation}
 R_{\mathrm{PD}}(\mathcal E_{\mathrm{BB84}})=1.
 \label{eq:BB84-optimized-PD}
\end{equation}

On the other hand, the uniform BB84 ensemble is the $d=2$ instance of the
uniform ensemble $\mathcal E_d$ formed from the computational and Hadamard
bases. Its accessible information satisfies \cite{divincenzo2004locking}
\begin{equation}
 I_{\mathrm{acc}}(\mathcal E_d)
 \leq \frac{1}{2}\log_2 d.
\end{equation}
Setting $d=2$ gives
$I_{\mathrm{acc}}(\mathcal E_{\mathrm{BB84}})\leq 1/2$.
Combining this upper bound with the achievable value in
Eq.~\eqref{eq:BB84-mutual-information}, we conclude that
\begin{equation}
 I_{\mathrm{acc}}(\mathcal E_{\mathrm{BB84}})
 =\frac{1}{2}.
\end{equation}
We therefore obtain the strict separation and Holevo saturation
\begin{equation}
 I_{\mathrm{acc}}(\mathcal E_{\mathrm{BB84}})
 =
 \frac12
 <
 R_{\mathrm{PD}}(\mathcal E_{\mathrm{BB84}})
 =
 \chi(\mathcal E_{\mathrm{BB84}})
 =
 1.
 \label{eq:BB84-separation}
\end{equation}

This equality does not imply that one bit of label information can be
decoded from an individual BB84 signal. Rather, \(R_{\mathrm{PD}}\) quantifies
the total information about \(A\) represented across the two
posterior-orthogonal record contexts.

\section{Proof of Theorem~\ref{thm:main-hierarchy-sec3}}
\label{sec:proof-main}

The proof of the upper bound relies on a recovery inequality for conditional mutual information.
We first introduce the channel formulation of a measurement and the recovery
result needed below.

A POVM $\mathsf M=\{E_b\}_{b\in\mathcal B}$ defines the
quantum-to-classical channel
\begin{equation}
 \mathcal M(\xi)
 =\sum_{b\in\mathcal B}\operatorname{Tr}(E_b\xi)
 |b\rangle\!\langle b|^B,
 \label{eq:measurement-channel-proof}
\end{equation}
where $\{|b\rangle^B\}_{b\in\mathcal B}$ is an orthonormal basis whose
labels record the measurement outcomes.

For quantum states $\rho$ and $\sigma$, the measured relative
entropy is defined as \cite{brandao2015quantum}
\begin{equation}
D_{\mathsf M}(\rho\Vert\sigma)
:=
\sup_{\mathcal M}D\!\left(\mathcal M(\rho)\Vert\mathcal M(\sigma)\right),
\end{equation}
where the supremum is over all finite-outcome
measurement channels $\mathcal M$. For every quantum channel
$\Lambda$, it satisfies the data-processing inequality
\begin{equation}
D_{\mathsf M}(\rho\Vert\sigma)
\geq
D_{\mathsf M}\!\left(\Lambda(\rho)\Vert\Lambda(\sigma)\right).
\end{equation}
This follows directly from the
definition, since $\mathcal M\circ\Lambda$ is itself a measurement
channel.

\begin{lemma}[Measured-relative-entropy recovery]
\label{lem:measured-recovery}
For every finite-dimensional tripartite state $\omega^{ABC}$,
\begin{equation}
 I(A:C|B)_\omega
 \geq
 \inf_{\mathcal R_{B\to BC}}
 D_{\mathrm M}\!\left(
 \omega^{ABC}\middle\|
 (\operatorname{id}_A\otimes\mathcal R)(\omega^{AB})
 \right),
 \label{eq:recovery-inequality}
\end{equation}
where the infimum is over CPTP recovery maps
$\mathcal R_{B\to BC}$.
\end{lemma}

Lemma~\ref{lem:measured-recovery} is the single-copy
measured-relative-entropy recovery bound of
Ref.~\cite{brandao2015quantum}. Here, for a fixed recovery map $\mathcal R_{B\to BC}$, the recovery
error is the measured relative entropy between the original state
$\omega_{ABC}$ and the recovered state
$(\id_A\otimes\mathcal R_{B\to BC})(\omega_{AB})$. The lemma states
that the infimum of this error over all recovery maps is upper-bounded
by the conditional mutual information.

We next connect the recovery lemma to the directed posterior novelty.
\begin{lemma}[Directed posterior-novelty bound]
\label{lem:directed-novelty}
Let $\mathcal M_i:H\to B_i$ and $\mathcal M_j:H\to B_j$ be
finite-outcome measurement channels for the finite-dimensional cq state
$\rho^{AH}$. If $i\neq j$, then
\begin{equation}
 \delta_{j\to i}
 \leq I(A:H)_\rho-I(A:B_i)_{\rho^{AB_i}},
 \label{eq:directed-novelty-bound}
\end{equation}
where
$\rho^{AB_i}=(\operatorname{id}_A\otimes\mathcal M_i)(\rho^{AH})$.
\end{lemma}

\begin{proof}
Choose a Stinespring isometry $V_i:H\to B_iE_i$ for $\mathcal M_i$, so that
\begin{equation}
 \mathcal M_i(\xi)=\operatorname{Tr}_{E_i}(V_i\xi V_i^\dagger),
\end{equation}
and define
\begin{equation}
 \omega^{AB_iE_i}
 =(\operatorname{id}_A\otimes V_i)\rho^{AH}
  (\operatorname{id}_A\otimes V_i^\dagger).
 \label{eq:dilated-state}
\end{equation}
Isometric invariance and the chain rule imply
\begin{align}
 I(A:H)_\rho-I(A;B_i)_\omega
 &=I(A:B_iE_i)_\omega-I(A;B_i)_\omega\nonumber\\
 &=I(A:E_i|B_i)_\omega.
 \label{eq:loss-as-cmi}
\end{align}
Thus the right-hand side of Eq.~\eqref{eq:directed-novelty-bound} is both the
mutual information lost under measurement $i$ and a conditional mutual
information in its dilation.

By Eq.~\eqref{eq:recovery-inequality}, for every $\varepsilon>0$ there is a
CPTP map $\mathcal R_i:B_i\to B_iE_i$ such that, with
\begin{equation}
 \widetilde\omega^{AB_iE_i}
 =(\operatorname{id}_A\otimes\mathcal R_i)(\omega^{AB_i}),
\end{equation}
one has
\begin{equation}
 I(A:E_i|B_i)_\omega+\varepsilon
 \geq D_{\mathrm M}(\omega\|\widetilde\omega).
 \label{eq:recovered-state-bound}
\end{equation}

To compare this recovery error with \(\delta_{j\to i}\), map the recovered
system back to \(H\). Let \(P_i=V_iV_i^\dagger\), choose an arbitrary density
operator \(\tau_0\) on \(H\), and define
\begin{equation}
 \mathcal D_i(Z)
 :=V_i^\dagger ZV_i
  +\operatorname{Tr}\!\left[(\mathbb I-P_i)Z\right]\tau_0 .
 \label{eq:left-inverse}
\end{equation}
The map \(\mathcal D_i:B_iE_i\to H\) is CPTP and satisfies
\[
 \mathcal D_i(V_i\xi V_i^\dagger)=\xi .
\]
Apply the channel
\begin{equation}
 \Phi_j
 :=\operatorname{id}_A\otimes
   (\mathcal M_j\circ\mathcal D_i)
\end{equation}
to both states in Eq.~\eqref{eq:recovered-state-bound}. The channel \(\Phi_j\) maps the true state \(\omega^{AB_iE_i}\) to the
classical state corresponding to \(p_j(A,B_j)\):
\begin{equation}
 \Phi_j(\omega^{AB_iE_i})
 =
 \sum_{a,b_j}p_j(a,b_j)
 |a\rangle\!\langle a|
 \otimes |b_j\rangle\!\langle b_j| .
 \label{eq:true-output-statistics}
\end{equation}
Let \(q(A,B_j)\) denote the statistics generated by applying the same channel
to the recovered state
\[
 \widetilde\omega^{AB_iE_i}
 =
 (\operatorname{id}_A\otimes\mathcal R_i)(\omega^{AB_i}).
\]
Thus,
\begin{equation}
 \Phi_j(\widetilde\omega^{AB_iE_i})
 =
 \sum_{a,b_j}q(a,b_j)
 |a\rangle\!\langle a|
 \otimes |b_j\rangle\!\langle b_j| .
 \label{eq:recovered-output-statistics}
\end{equation}
Because \(B_i\) is classical, its marginal has the form
\begin{equation}
 \omega^{AB_i}
 =
 \sum_{a,b_i}p_i(a,b_i)
 |a\rangle\!\langle a|
 \otimes |b_i\rangle\!\langle b_i| .
 \label{eq:classical-measurement-state}
\end{equation}
For the recovery map \(\mathcal R_i\) chosen above, define
\begin{equation}
 \sigma_{b_i}^{B_iE_i}
 :=
 \mathcal R_i\!\left(|b_i\rangle\!\langle b_i|\right)
\end{equation}
and
\begin{equation}
 K(b_j|b_i)
 :=
 \left\langle b_j\middle|
 (\mathcal M_j\circ\mathcal D_i)(\sigma_{b_i})
 \middle|b_j\right\rangle .
 \label{eq:induced-stochastic-map}
\end{equation}
For every \(b_i\), \(K(\cdot|b_i)\) is a probability distribution. Indeed,
\(\sigma_{b_i}\) is a density operator and
\(\mathcal M_j\circ\mathcal D_i\) is a measurement channel.

Using Eq.~\eqref{eq:classical-measurement-state}, the recovered state can be
written as
\begin{equation}
 \widetilde\omega^{AB_iE_i}
 =
 \sum_{a,b_i}p_i(a,b_i)
 |a\rangle\!\langle a|\otimes\sigma_{b_i}^{B_iE_i}.
 \label{eq:recovered-state-expansion}
\end{equation}
Applying \(\Phi_j\) to this expression and using linearity gives
\begin{equation}
 q(a,b_j)
 =
 \sum_{b_i}p_i(a,b_i)K(b_j|b_i).
 \label{eq:recovered-distribution}
\end{equation}
Hence, whenever \(q(b_j)>0\),
\begin{equation}
 q(A|b_j)
 =
 \sum_{b_i}
 \frac{p_i(b_i)K(b_j|b_i)}{q(b_j)}
 p_i(A|b_i)
 \in\mathcal C_i .
 \label{eq:recovered-posterior-in-hull}
\end{equation}
Thus the statistics generated from \(\widetilde\omega\) are obtained by
stochastic post-processing of the record \(B_i\). 

Data processing of measured relative entropy and classicality of the
outputs now give
\begin{align}
 D_{\mathrm M}(\omega\|\widetilde\omega)
 &\geq D\!\left(p_j(A,B_j)\middle\|q(A,B_j)\right)\nonumber\\
 &=D\!\left(p_j(B_j)\middle\|q(B_j)\right)\nonumber\\
 &\quad+\sum_{b_j}p_j(b_j)
 D\!\left(p_j(A|b_j)\middle\|q(A|b_j)\right)\nonumber\\
 &\geq\sum_{b_j}p_j(b_j)
 \inf_{r\in\mathcal C_i}
 D\!\left(p_j(A|b_j)\middle\|r\right)\nonumber\\
 &=\delta_{j\to i}.
 \label{eq:recovery-to-novelty}
\end{align}

If $p_j(b_j)>0$ but $q(b_j)=0$, the joint relative entropy is infinite,
contradicting the finite upper bound in Eq.~\eqref{eq:recovered-state-bound};
thus all relevant conditional distributions above are well defined.
Combining Eqs.~\eqref{eq:loss-as-cmi},
\eqref{eq:recovered-state-bound}, and \eqref{eq:recovery-to-novelty}, and
then letting $\varepsilon\rightarrow 0^{+}$, proves the claim.
\end{proof}

The lemma links posterior geometry to measurement loss: the novelty of the
posterior structure generated by $B_j$ relative to $\mathcal C_i$ cannot
exceed the correlations lost in measurement $i$.

Now we give the complete proof of Theorem \ref{thm:main-hierarchy-sec3}.

\begin{proof}[Proof of Theorem \ref{thm:main-hierarchy-sec3}]
 The proof of the lower bound $I_{\mathrm{acc}}(\mathcal E)
 \leq R_{\mathrm{PD}}(\mathcal E)$ is  immediate using the first inequality of Eq.~\eqref{rangeHSP} and the definition Eq.~\eqref{DefBySup}. 
 
For the upper bound $R_{\mathrm{PD}}(\mathcal E)
 \leq \chi(\mathcal E)$, 
 apply Lemma~\ref{lem:directed-novelty} in both
directions:
\begin{align}
 I(A;B_1)+\delta_{2\to1}&\leq I(A:H)_\rho,\nonumber\\
 I(A;B_2)+\delta_{1\to2}&\leq I(A:H)_\rho.
 \label{eq:two-directed-branches}
\end{align}
Taking the maximum of the two branches gives, for every pair of POVMs,
\begin{equation}
 R_{\mathrm{PD}}(\mathcal E;\mathcal M_1,\mathcal M_2)
 \leq I(A:H)_\rho
 =\chi(\mathcal E).
 \label{eq:fixed-pair-holevo-bound}
\end{equation}
Because $\chi(\mathcal E)$ is a common upper bound for all fixed-pair values,
taking their supremum preserves the inequality:
\begin{equation}
 R_{\mathrm{PD}}(\mathcal E)
 \leq\chi(\mathcal E).
 \label{eq:optimized-holevo-bound}
\end{equation}  
\end{proof}

\section{Quantum validity of the redundancy-based information causality condition}
\label{sec:bounding-quantum-correlations}

The hierarchy established above provides a quantum-validity
proof for the redundancy-based information causality (IC) condition
proposed in Ref.~\cite{yu2026information}. We first recall the relevant
IC scenario and the definition of this
condition.

The IC scenario provides a candidate operational bound.
Alice receives a classical variable $A$ and communicates with Bob through a
classical channel of capacity $k$, while they may share a nonsignalling
resource established independently of $A$ \cite{pawlowski2009information,barnum2010entropy,pawlowski2015information,miklin2021information}.  After receiving the channel
output, Bob selects a decoding setting $t$ and produces a record $B_t$.
For a fixed encoding, communication channel, and decoding
scheme, the IC protocol can be simply represented as a mapping from a shared resource correlation to a family of observable
pair distributions,
\begin{equation}
 \mathcal P_{\mathrm{IC}}:
 P(a_r,b_r|x,y)
 \longmapsto
 \bigl\{p(A,B_t)\bigr\}_{t=1}^{N},
 \label{eq:ic-protocol-map}
\end{equation}
where $a_r$ and $b_r$ denote the outputs of the shared resource.  The pair
distributions on the right correspond to alternative decoding choices and
are not assumed to arise as marginals of a joint distribution of all
$B_t$.
The principle of IC postulates that the potential information about
Alice's input that arises through the communication and is quantified from the pair distributions
cannot exceed the channel capacity $k$ for any physical resource.

% In the original random-access-code (RAC) formulation of IC, Alice's input is imposed to be a string
% $A=(A_1,\ldots,A_N)$ of independent bits, and the setting $t=i$ asks Bob to
% guess the designated bit $A_i$. 
% The potential
% information is quantified by
% \begin{equation}
%  IC_{\mathrm{RAC}}
%  =\sum_{i=1}^{N} I(A_i;B_i),
% \end{equation}
% and information causality requires
% $IC_{\mathrm{RAC}}\leq k$.  
% This formulation establishes a
% componentwise correspondence between Alice's input bits and Bob's decoding
% settings: each term measures the information obtained about one designated
% bit, and their sum quantifies the information potentially accessible through
% the family of alternative decoding tasks.

Recently, a formulation of IC based on the HSP measure of redundancy was
proposed for two decoding settings \cite{yu2026information}.  Unlike the
original RAC formulation, which sums information about componentwise
designated input bits \cite{pawlowski2009information}, this formulation
treats $A$ as a single random variable.  Since the records generated by the
two decoding settings may represent overlapping information about $A$, the
HSP redundancy is subtracted when defining the potential information:
\begin{equation}
 IC_{\mathrm{red}}
 :=I(A;B_1)+I(A;B_2)
   -I_{\cap}^{\mathrm{HSP}}(A;B_1,B_2).
 \label{eq:ic-red-definition}
\end{equation}
The associated redundancy-based IC condition
is
\begin{equation}
 IC_{\mathrm{red}}\leq k.
 \label{ICredPrinciple}
\end{equation}
By Eq.~\eqref{eq:total-max-form}, the potential information
$IC_{\mathrm{red}}$ is exactly the fixed-pair PD total information
$R_{\mathrm{PD}}(A;B_1,B_2)$ in this paper. 
Ref.~\cite{yu2026information} showed numerically that
condition \eqref{ICredPrinciple} rules out beyond-quantum correlations missed
by the RAC criterion and reproduces the quantum boundary in some studied
slices,
but it did not prove that all quantum correlations satisfy the condition. 
Consequently, a violation of the condition could not, by itself,
be interpreted as a conclusive witness of beyond-quantum
correlations. 

We now use the hierarchy \eqref{MainHierarchy} to establish
the missing quantum-validity result under the protocol
assumptions stated above.  
For a quantum realization, let $H$ denote Bob's local quantum system
before receiving Alice's message, and let $M$ denote the classical
message he receives through communication. 
Conditioned on Alice's
label $A=a$, 
these systems define an ensemble
$\mathcal E=\{p_a,\rho_{HM}^a\}_a$ and the associated cq state
\begin{equation}
\rho_{AHM}
=
\sum_a p_a\,\ket{a}\!\bra{a}_A\otimes\rho_{HM}^a .
\end{equation}
Because Alice's input is independent of the initially shared resource,
and because the resource is nonsignalling, Bob's precommunication
system satisfies $I(A:H)_\rho=0$. Hence
\begin{align}
\chi(\mathcal E)
&= I(A:HM)_\rho \nonumber
 \leq k ,
\label{eq:ic-holevo-capacity-bound}
\end{align}
where the inequality is proved in Ref.~\cite{pawlowski2015information}.
Combining this bound with the
HSP--Holevo hierarchy yields, for any two decoding quantum measurements,
\begin{equation}
 IC_{\mathrm{red}}
 =R_{\mathrm{PD}}(A;B_1,B_2)
 \leq R_{\mathrm{PD}}(\mathcal E)
 \leq\chi(\mathcal E)
 \leq k.
 \label{eq:analytic-ic-red-bound}
\end{equation}
Thus, under the protocol assumptions stated above,
condition~\eqref{ICredPrinciple} is satisfied by every quantum
realization, establishing it as a quantum-valid formulation of IC. 
If a correlation
$P(a_{\mathrm r},b_{\mathrm r}|x,y)$ generates pair distributions that violate
condition~\eqref{ICredPrinciple}, then no quantum realization of that
correlation is compatible with the protocol. Such a violation therefore
certifies beyond-quantum correlations within this setting.

\section{Discussion and conclusion}
\label{sec:discussion}

We have defined the PD total information represented across two alternative
measurement contexts using only their pair distributions.
For posterior-orthogonal records, this quantity reduces to the sum of the two
record mutual informations.
For every
finite-dimensional cq ensemble, its optimization lies between the accessible
and Holevo information,
\begin{equation}
 I_{\mathrm{acc}}(\mathcal E)
 \leq R_{\mathrm{PD}}(\mathcal E)
 \leq \chi(\mathcal E).
 \label{eq:conclusion-hierarchy}
\end{equation}
The upper bound follows by relating directed posterior novelty to
measurement-induced information loss through quantum recovery.  
The uniform BB84 ensemble exhibits strict accessible-information separation and Holevo
saturation. Finally, applying the hierarchy to the previously proposed
HSP redundancy-based information causality condition establishes
its validity for quantum realizations under the corresponding
protocol assumptions. 

Several questions remain open. For posterior-orthogonal records,
\(R_{\mathrm{PD}}\) equals the sum of the two record mutual informations,
reflecting their vanishing HSP projected redundancy. In general, however,
\(R_{\mathrm{PD}}\) remains a geometric quantity defined through
relative-entropy projections. Finding a direct operational interpretation,
or a protocol whose value is characterized by \(R_{\mathrm{PD}}\), remains
an important open problem.

A second challenge is extension beyond two decoding settings.  
For three or more settings, 
a direct extension would require a
consistent treatment of overlaps among several posterior families, which may
involve a joint coupling of the alternative records.  Such a joint structure
is generally unavailable for incompatible measurements and is not determined
by the observed pair distributions. Whether a genuinely multivariate redundancy can be defined from the family $\{p(A,B_t)\}_{t=1}^{N}$ alone, is therefore an important
open problem.

Finally, recent work has shown that there exist post-quantum correlations
that satisfy all currently known formulations of information causality,
including the HSP redundancy-based formulation
\cite{pompili2026information}.  Thus, although a violation of the HSP
information causality condition certifies beyond-quantum correlations, its
satisfaction does not establish quantumness.  
It is therefore natural to ask
whether suitably designed protocols, or alternative notions of total
information derived from partial information decomposition, can recover the
quantum boundary in specific correlation scenarios.

\section*{acknowledgement}
The authors thank Valerio Scarani, Masahito Hayashi, Dong Yang, and
Francesco Buscemi for helpful discussions during the early stages of this
work. B.Y. was supported by the GuangDong Basic and Applied Basic Research Foundation (Grant No. 2024A1515012405). 
The authors used
ScienceLab AutoProject to assist with the development of the proof and the
presentation of the manuscript. In particular, it suggested the use of
recovery-channel techniques in the proof of
Theorem~\ref{thm:main-hierarchy-sec3}. 
The authors reviewed and verified all
arguments and take full responsibility for the content of the manuscript. 

\section*{Data Availability}
No data were created or analyzed in this study.

\bibliography{ref}

\appendix
\section{Proof of Proposition~2}\label{app:posterior-orthogonality}

We prove that posterior orthogonality makes the prior an information
projection of every posterior generated by one record onto the posterior
polytope generated by the other.  Fix $i\neq j$ and an outcome $b_j$ with
$p_j(b_j)>0$, and write
\begin{equation}
 q(a):=p_j(a|b_j).
\end{equation}
For any $r\in\mathcal C_i$, there are coefficients $\lambda(b_i)\geq0$ with
$\sum_{b_i}\lambda(b_i)=1$ such that
\begin{equation}
 r(a)=\sum_{b_i}\lambda(b_i)p_i(a|b_i).
 \label{eq:app-r-mixture}
\end{equation}
Set $d(a):=r(a)-p(a)$.  By Eq.~\eqref{eq:app-r-mixture},
\begin{equation}
 d(a)=\sum_{b_i}\lambda(b_i)
       \bigl[p_i(a|b_i)-p(a)\bigr].
 \label{eq:app-displacement}
\end{equation}
Posterior orthogonality, Eq.~(15), and
linearity therefore give
\begin{equation}
 \sum_a\frac{[q(a)-p(a)]d(a)}{p(a)}=0.
 \label{eq:app-orthogonality}
\end{equation}
Moreover, $\sum_a d(a)=0$ because $r$ and $p$ are normalized.  Hence
\begin{equation}
 \sum_a\frac{q(a)d(a)}{p(a)}
 =\sum_a\frac{[q(a)-p(a)]d(a)}{p(a)}+\sum_a d(a)
 =0.
 \label{eq:app-stationarity}
\end{equation}

Consider the line segment $r_t=p+t(r-p)=p+td$, with $0\leq t\leq1$, and
define
\begin{equation}
 f(t):=D(q\|r_t).
\end{equation}
Relative entropy is convex in its second argument, so $f$ is convex.  Since
$p(a)>0$ on its support, its right derivative at $t=0$ is well defined and,
for base-two logarithms, satisfies
\begin{equation}
 f'(0^+)=-\frac{1}{\ln 2}
 \sum_a\frac{q(a)d(a)}{p(a)}=0,
 \label{eq:app-derivative}
\end{equation}
where Eq.~\eqref{eq:app-stationarity} was used.  Convexity then implies
\begin{equation}
 D(q\|r)=f(1)\geq f(0)=D(q\|p).
 \label{eq:app-prior-minimizer}
\end{equation}
The inequality also holds when $D(q\|r)=+\infty$.  Because
$p(A)\in\mathcal C_i$, equality is attained by choosing $r=p$.  Consequently,
\begin{equation}
 \inf_{r\in\mathcal C_i}
 D\!\left(p_j(A|b_j)\middle\|r\right)
 =D\!\left(p_j(A|b_j)\middle\|p(A)\right).
 \label{eq:app-information-projection}
\end{equation}

Averaging Eq.~\eqref{eq:app-information-projection} over $b_j$ yields
\begin{align}
 \delta_{j\to i}
 &=\sum_{b_j}p_j(b_j)
 D\!\left(p_j(A|b_j)\middle\|p(A)\right)\notag\\
 &=I(A;B_j),
 \qquad i\neq j.
 \label{eq:app-full-novelty}
\end{align}
Thus the directed novelty of either record relative to the other equals its
full mutual information with $A$.  Substituting
$\delta_{2\to1}=I(A;B_2)$ and $\delta_{1\to2}=I(A;B_1)$ into the definition
of $R_{\mathrm{PD}}$ gives
\begin{align}
 R_{\mathrm{PD}}(A;B_1,B_2)
 &=\max\!\left\{
 I(A;B_1)+\delta_{2\to1},
 I(A;B_2)+\delta_{1\to2}
 \right\}\notag\\
 &=I(A;B_1)+I(A;B_2),
\end{align}
which proves Proposition~2.

\end{document}